# State of the Practice in Software Effort Estimation: A Survey and Literature Review

Adam Trendowicz[1], Jürgen Münch[1], Ross Jeffery[2, 3]

[1]Fraunhofer IESE, Fraunhofer-Platz 1, 67663 Kaiserslautern, Germany
{trend, muench}@iese.fraunhofer.de

[2]University of New South Wales, School of Computer Science and Engineering
Sydney 2052, Australia

[3]National ICT Australia, Australian Technology Park, Bay 15 Locomotive Workshop,
Eveleigh NSW 2015, Australia
rossj@cse.unsw.edu.au

**Abstract.** Effort estimation is a key factor for software project success, defined as delivering software of agreed quality and functionality within schedule and budget. Traditionally, effort estimation has been used for planning and tracking project resources. Effort estimation methods founded on those goals typically focus on providing exact estimates and usually do not support objectives that have recently become important within the software industry, such as systematic and reliable analysis of causal effort dependencies. This article presents the results of a study of software effort estimation from an industrial perspective. The study surveys industrial objectives, the abilities of software organizations to apply certain estimation methods, and actually applied practices of software effort estimation. Finally, requirements for effort estimation methods identified in the survey are compared against existing estimation methods.

## 1 Introduction

Rapid growth in the demand for high-quality software and increased investments in software projects show that software development is one of the key markets worldwide. The average company spends about 4 to 5 percent of its revenues on information technology, with those that are highly IT-dependent - such as financial and telecommunications companies - spending more than 10 percent [6]. A fast changing market demands software products with ever more functionality, higher reliability, and higher performance. In addition, in order to stay competitive, senior managers increasingly demand that IT departments deliver more functionality and quality with fewer resources, and do it faster than ever before [7]. Together with the increased complexity of software, the global trend towards shifting development from single contractors to distributed projects has led to software companies needing a reliable basis for making make-or-buy decisions. Finally, software development teams must strive to achieve all these objectives by exploiting the impressive advances in continuously changing (and often immature) technologies.

In this situation, software planning and management seem to be an essential and, at the same time very difficult task. Dozens of well-known software disasters [6] are the most visible examples of problems in managing complex, distributed software systems. Independent of such experiences, many software organizations are still proposing unrealistic software costs, work within tight schedules, and finish their projects behind schedule and budget (46%), or do not complete them at all (19%) [19]. Finally, even if completed within a target plan, overestimated projects typically expand to consume whatever more resources were planned, while the functionality and quality of underestimated projects is cut to fit the plan.

To address these and many other issues, considerable research has been directed at building and evaluating software effort estimation methods and tools [5, 20]. For decades, estimation accuracy was a dominant prerequisite for accepting or declining a certain effort estimation method. Yet, even though perhaps very accurate, it does not guarantee project success. Software decision makers have recently had to face this problem and find an estimation method that, first, is *applicable* in the context of a certain organization and, second, *contributes* to the achievement of organization-specific objectives. Therefore, in order to accept or decline a certain estimation method, two major criteria should be met. The *necessary acceptance criterion* is that the software organization is able to fulfill the application prerequisites of a certain method, such as needed measurement data or required involvement of human experts. The *sufficient acceptance criterion*, on the other hand, refers to the method's contribution to organizational objectives. As reported in this paper, besides accuracy, estimation methods are expected to support a variety of project management activities, such as risk analysis, benchmarking, or process improvement.

In this article, we survey industrial objectives with respect to effort estimation, the abilities of software organizations to apply certain estimation methods, and effort estimation practices actually applied within the software industry. The remainder of the paper is organized as follows: Section 2 defines common sources of effort estimation problems. Section 3 outlines the study design. Section 4 summarizes current industrial practices with respect to software effort estimation, followed (Section 5) by a comparative evaluation of existing effort estimation methods concerning their suitability to meet certain industrial requirements. Finally, Section 6 concludes the results of the study presented in this paper and provides further research directions.

## 2 Sources of Deficits in Software Effort Estimation

As with any other process or technology, software effort estimation is expected to meet certain objectives and requires certain prerequisites to be applied. A problem occurs where the applied estimation method is not in line with the defined objectives and/or the available resources. Specifically, we identify two major sources of effort estimation problems in the software industry, which may be addressed by the following questions:

- *What is provided by a method?* The estimation method provides less than required for the achievement of organizational objectives, e.g., it provides only a point estimate and thus hardly any support for managing project risks if the estimate is lower than the available budget.
- *What is required by a method?* The estimation method requires more resources than currently available in the organization, e.g., it requires more measurement data than actually available in the organization.

## 3 Study Design

The study presented in this paper consisted of two elements: (1) a review of related literature and (2) a survey performed among several software companies.

### 3.1 Study Objectives

The objective of the study was the analysis of the current industrial situation regarding software effort estimation practices (*state of the practice*). In particular, the following aspects were focused on:

- *Effort estimation objectives*: industrial objectives (expectations) with respect to effort estimation methods;
- *Effort estimation abilities*: ability of software companies to apply a certain estimation approach, e.g., by providing necessary resources;
- *Effort estimation methods*: effort estimation methods actually applied at software companies.

In addition, existing estimation methods are analyzed concerning how they meet industrial needs and abilities regarding software effort estimation (*state of the art*).

### 3.2 Information Sources

The study presented in this paper is based on two sources of information:

- *Literature Revi*ew: Literature published in the public domain such as books, journals, conference proceedings, reports, and dissertations.
- *Industrial Surveys*: Information gained during two series of industrial surveys (S1 and S2) performed at 10 software organizations. The survey results presented in this chapter include both surveys, unless explicitly stated otherwise.

## 3.3 Literature Review

The design of the review is based on the guidelines for systematic reviews in software engineering proposed, for instance, in [4, 8]. It was performed as follows (Fig. 1):

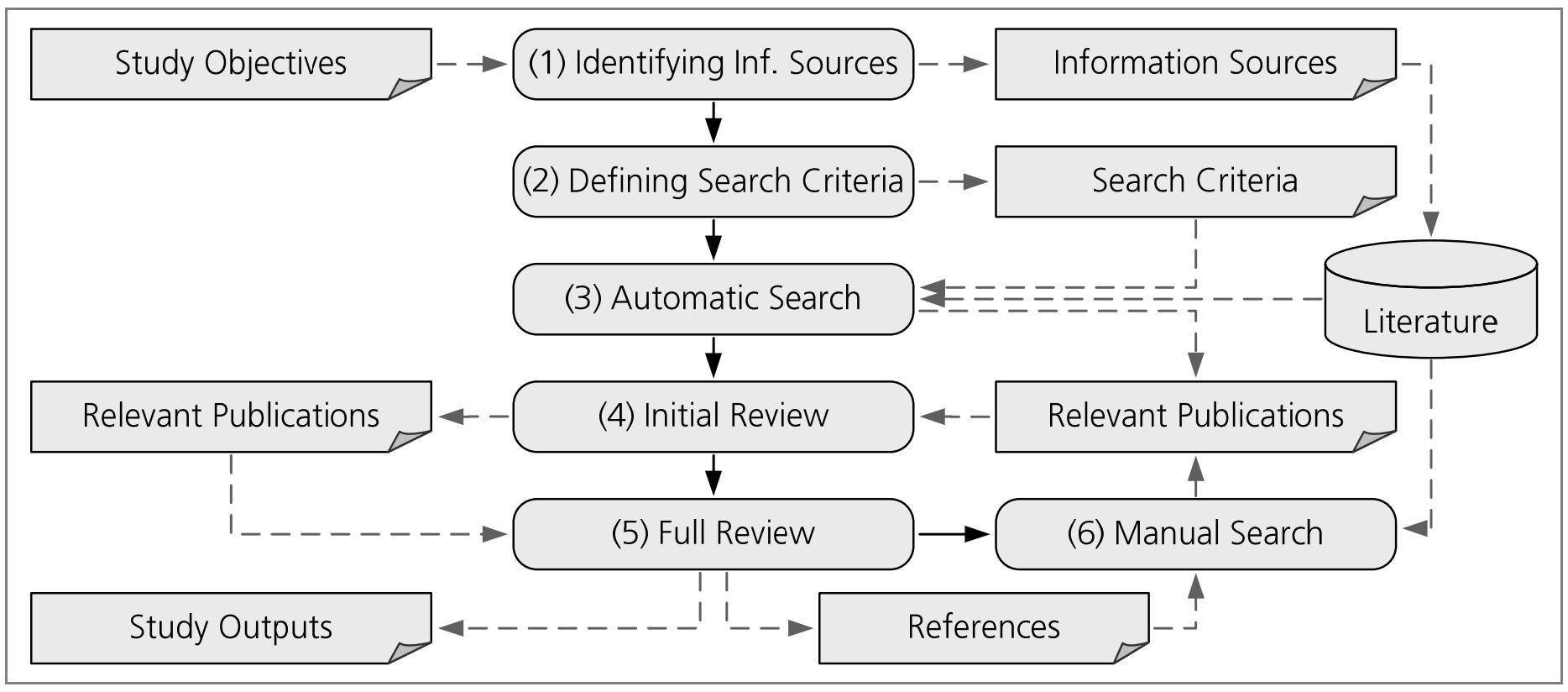

**Fig. 1** Overview of the literature review process

1. ***Identifying information sources***: We identified the most relevant sources of information with respect to the study objectives. Initially, they included the most prominent software engineering journals and conference proceedings[1].
2. ***Defining search criteria***: We limited the review scope by two criteria: relevancy and recency. *Relevancy* defined the content of the reviewed publications. We focused on the titles that dealt with software effort estimation and related topics such as development productivity. *Recency* defined the time frame of the reviewed publications. Since software engineering is one of the most rapidly changing domains, we decided to limit our review to papers published after the year 1995. The analysis of industrial practices encompassed an even narrower scope, namely publication after the year 2000.
3. ***Automatic search***: We performed an automatic search through relevant sources of information using defined search criteria (Table 1). This search included a search for specific keywords in the publications' title, abstract, and list of keywords (if provided). We used generic search engines such as INSPEC (http://www.iee.org/publish/inspec/) and specific engines associated with concrete publishers such as IEEE Xplore (http://ieeexplore.ieee.org/Xplore/dynhome.jsp).
4. ***Initial review***: We initially reviewed the title, abstract, and keywords of the publications returned by the automatic search with respect to the defined criteria for inclusion in the final review.
5. ***Full review***: A complete review of the publications accepted during the initial review was performed. This step was followed by a manual search and review (if accepted) of referenced relevant publications that had not been found in earlier steps of the review. In total, 380 publications were reviewed in the study.

[1] For a detailed list of the information sources considered in the study, please refer to [20].

6. ***Manual search***: In accordance with the recommendations presented in [8], we complemented the results of the automatic search with relevant information sources by doing a manual search through references found in reviewed papers as well as using a generic web search engine (http://www.google.com).

**Table 1.** Query defined for the purpose of automatic search

| |
|---|
| ((Software OR Project) AND (Effort OR Cost OR Estimation OR Prediction) IN (Title OR Abstract OR Keywords)) AND (Date >= 1995 AND Date <= 2008) |

### 3.4 Industrial Surveys

During the years 2005-2008, we performed a series of industrial surveys aimed at analyzing current industrial practices with respect to modeling (estimation and measurement) software development effort and productivity. In this paper, we present the aggregated results of the two most recent surveys and one industry workshop.

The early survey (S1) focused on effort estimation practices as well as closely related productivity and size measurement practices. The survey encompassed 7 software organizations. The recent survey (S2) focused specifically on effort estimation practices. It included such particular issues as objectives of estimation, estimation process, as well as inputs and outputs of estimation. The survey was performed in 2 software companies, with one of them being represented by 7 different development groups. In addition, we include the results of an effort estimation workshop we performed in 2007 in the software business unit of a large international provider of software systems (embedded domain). During the workshop, we asked the 7 participants about their estimation objectives, the estimation methods currently applied, and organizational capabilities with respect to effort modeling.

**Table 2.** Overview of the industrial survey characteristics

| Id | Size | Org. Type | Involved | App. Type | Domain |
|---|---|---|---|---|---|
| S1.1 | L | Supplier | DM, PM | Finance | MIS |
| S1.2 | L | Supplier | QM, QE, PM | Automation Systems | MIS |
| S1.3 | L | Supplier | PM, PM, QE | Finance | MIS |
| S1.4 | L | Supplier | PM, PM, QE | Finance | MIS |
| S1.5 | L | Supplier | PM | Finance | MIS |
| S1.6 | S | Supplier | DM, PM, PM | Finance | MIS |
| S1.7 | S | Supplier | PM, PM, QM | Finance | MIS |
| S2.1 | L | Supplier | 7 x PM | Medical, Automotive | EM |
| S2.2 | S | IV&V | QE | Space | EM |
| W1 | S | Supplier | DM, 2 x SD, 2 x QE, 2 x PM | Automation Systems | EM |

Table 2 presents an overview of the survey characteristics. For each survey, the following characteristics are included:

- *The size of the organization* considered in the survey in terms of number of employees. We distinguish Small (less than 50 employees), Medium (between 50 and 200 employees), and Large (more than 200 employees).

- *The type of involved organizations* was mainly software suppliers. In one case, the company provided independent verification and validation services (IV&V).
- *The involved roles* mainly included middle-level management and consultants: software developers (SD), project managers (PM), quality engineers (QE), quality managers (QM), and group/division managers (DM).
- *The application type* of typical products considered in the context of the involved organizations ranged from financial via medical, automotive, and industrial automation to safety-critical systems.
- *The domain* of the organization includes mainly information systems (MIS) and embedded software (EM). Information systems covered mainly financial systems, whereas embedded software covered such application areas as medical, automotive, telecommunication, and industrial automation.

### 3.5 Study Limitations

A major limitation of the study is the limited availability of information sources. The small sample of respondents in the industrial surveys might not be representative of the population of the software industry. Yet, since the results of the survey largely conform with the results of the literature review and our informal experiences gained during multiple industrial collaborations, we conclude that the results presented in this paper represent current trends in software effort estimation theory and practice.

## 4 State of the Industrial Software Effort Estimation Practice

The industrial practices surveyed in this paper include: (1) objectives of effort estimation, (2) capabilities to meet prerequisites for applying a certain estimation method, and (3) effort estimation methods currently applied in the software industry.

### Objectives of Effort Estimation

Fig. 2 presents a summary of effort estimation objectives identified in the study. In the case of several objectives, there is a noticeable discrepancy between what is presented in the related literature and what we observed during industrial surveys. From the perspective of study limitations, the availability of information sources may have resulted in data samples that are not representative of the whole software industry. In practice, there are, however, several other sources of variance in the results presented in Fig. 2. Traditional effort estimation objectives, such as planning and tracking the software project, or reducing project management overhead, are so common (and thus considered so "obvious") that they are typically not indicated if not asked for explicitly. Although the most recent literature focused most probably on the new objectives that have traditionally not been considered by the software industry until now, the survey also explicitly asked about traditional effort estimation objectives. Finally, we believe that the industrial surveys show some of the most recent trends that were not observed by the authors of the reviewed literature. Our personal experience confirms, for example, the increasing importance of process/productivity improvement objectives.

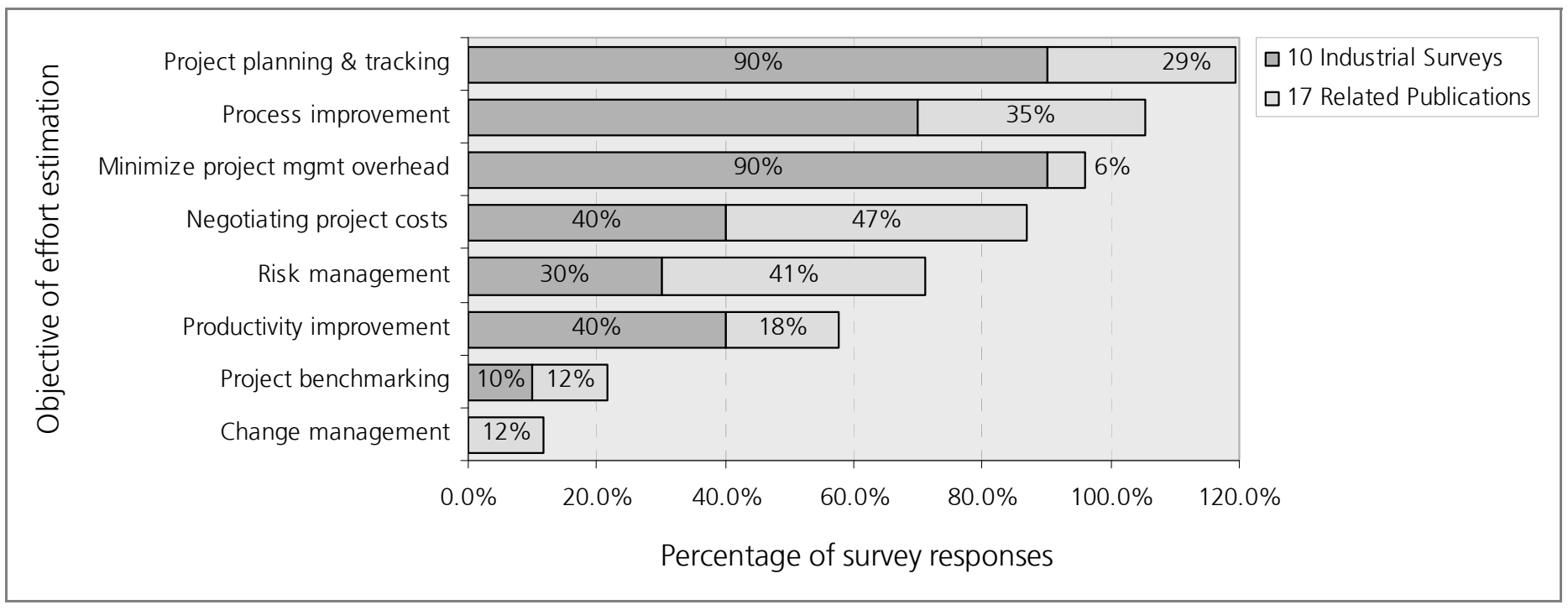

**Fig. 2.** Summary of objectives regarding software effort estimation

Based on the review of related literature and the results of the industrial surveys, we identified the 7 most common objectives regarding effort estimation methods:

1. ***Project planning & tracking***. The effort estimation method should support effective project planning by providing reliable and exact estimates at various levels of project granularity. There are two different ways of project planning dependent on the project type. In the context of *fixed-price* projects, effort estimation provides the basis for planning software functionality (how much functionality can be developed at a specified quality level and within a fixed budget). In the context of *fixed-product* projects, effort estimation provides a basis for planning software project schedule and cost (how much time and cost are required for developing software with a specified functionality and quality).
2. ***Process improvement***. The effort estimation method should support the understanding and improvement of effort- and productivity-related development processes.
3. ***Project management overhead***. The effort estimation method should minimize management overhead, i.e., the cost of applying and maintaining an estimation method (e.g., model building, application, and maintenance).
4. ***Negotiating project costs***. The effort estimation method should support the communication and negotiation process (justifying development costs) in the context of software procurement between the stakeholders involved (project managers, management, customers, etc.).
5. ***Risk management***. The effort estimation method should support management of project risks. This includes reciprocal integration of effort estimation and risk management, i.e., effort estimation uses the outputs of risk management (identified project risks having an impact on estimated effort) and provides input to risk management (cost-related risks). Moreover, the method should explicitly cover estimation uncertainty (i.e., accept uncertain/incomplete inputs and provide evaluation of the output's uncertainty).

6. ***Productivity improvement***. The effort estimation method should support the identification of factors that have the greatest influence on development productivity. Achievement of objective 1 usually implicates productivity improvement, at least from a long-term perspective; yet productivity might be improved without improving related processes, e.g., by assigning more skilled people to the project instead of improving training processes.
7. ***Project benchmarking***. The effort estimation method should support benchmarking of software projects with respect to development effort and productivity. Comparing software projects regarding productivity and effort between different organizations is especially important nowadays in the context of rapidly growing global development (out-sourcing, off-shoring, etc.) in order to support make-or-buy decisions and select/manage software suppliers.
8. ***Change management***. The effort estimation method should be easy to reapply along the software development life cycle in order to support managing changes in project scope, such as modified (non-)functional software requirements.

## 4.2 Effort Estimation Capabilities

The results of the industrial surveys and the literature review indicate that software organizations represent a very wide range of estimation capabilities.

***Estimation budget***: Due to extensive project pressure, software organizations do not typically assign sufficient manpower for estimation that is adequate to the approach applied. Depending on the specific domain, software organizations spend between 2% and 15% (on average 6%) of the entire project budget, which is typically not sufficient for applying estimation methods based on the judgment of multiple human experts.

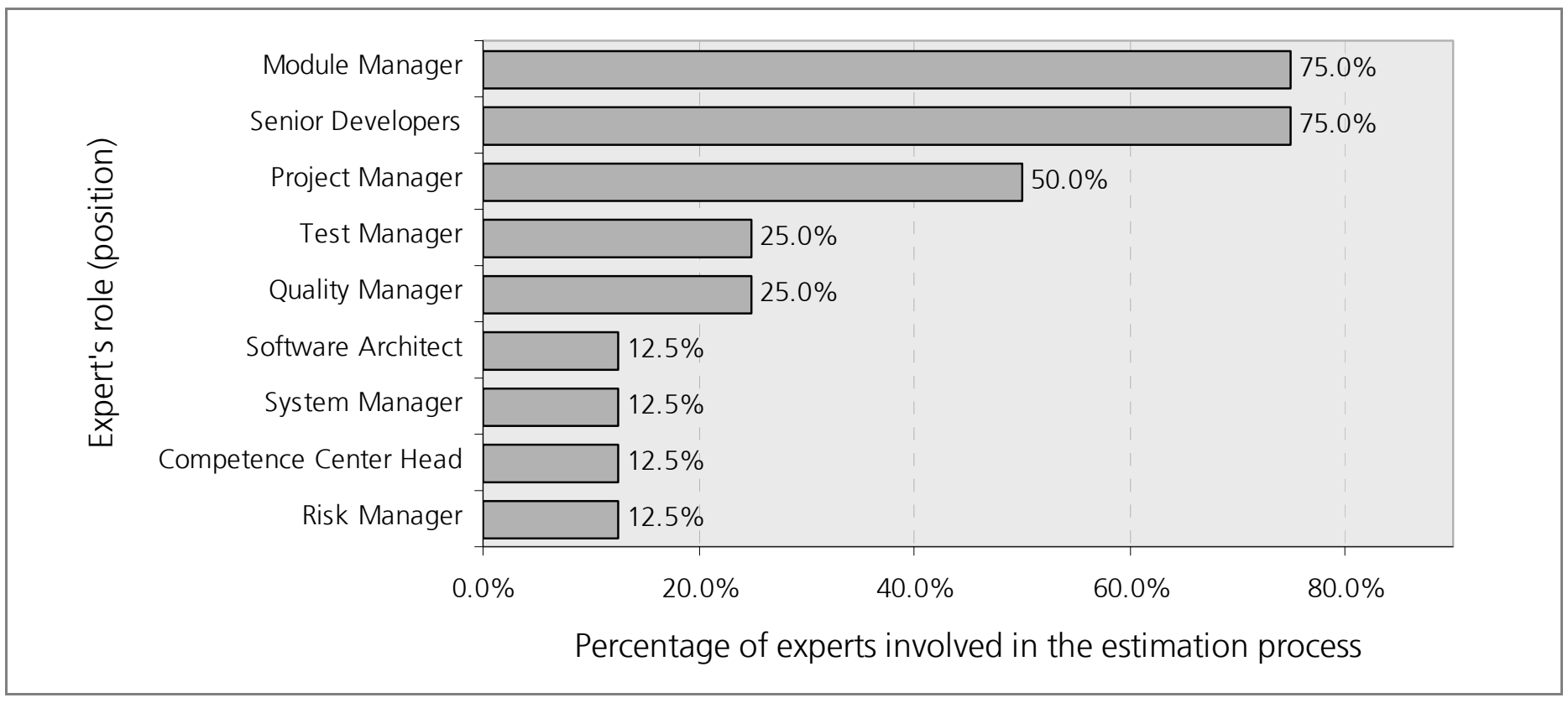

**Fig. 3.** Roles involved in the estimation process (survey S2)

***Expert involvement***: At the same time, respondents highlighted insufficient resources for estimation as one of the major problems that reduce the applicability and effectiveness of project effort prediction. In practice, software organizations are very unwilling to spend the effort of human experts on estimation tasks when they are responsible for other activities in a software project (Fig. 3). In consequence, estimation is typically performed by people lacking the appropriate expertise and/or with tight resources (e.g., insufficient budget designated for estimation).

***Quantitative project data***: On the other hand, software organizations typically do not have enough reliable project data to employ data-driven estimates and relieve human estimators. We observed (Fig. 4) that most of the time, the historical data a software organization can typically base estimates on does not exceed 10 projects. Even when the required amount of data has been collected, they are often not driven by specific objectives and often suffer from very low quality (inconsistency, incompleteness, etc.). The declared ratio of missing data ranges between 20% and 30%. Moreover, on average, 10-20% of the projects in the data repository can be described as exceptional projects that are not likely to be repeated (data outliers). One of the surveyed organizations, for example, collected measurement data from around 600 projects; yet, due to significant inconsistency and incompleteness, it was practically useless for estimation purposes (although great effort had been invested to validate and preprocess the data). This shows how important it is to define a proper, goal-oriented measurement program before data collection begins. Another interesting observation is that the quality and quantity of available data does not seem to be related to the specific reference process models used with an organization. A CMMI-L3 or -L5 organization may, for instance, have problems with collecting the minimal amount of useful project data, whereas an ISO9000 organization may already have a number of business-relevant measures collected for dozens of projects in their data repository.

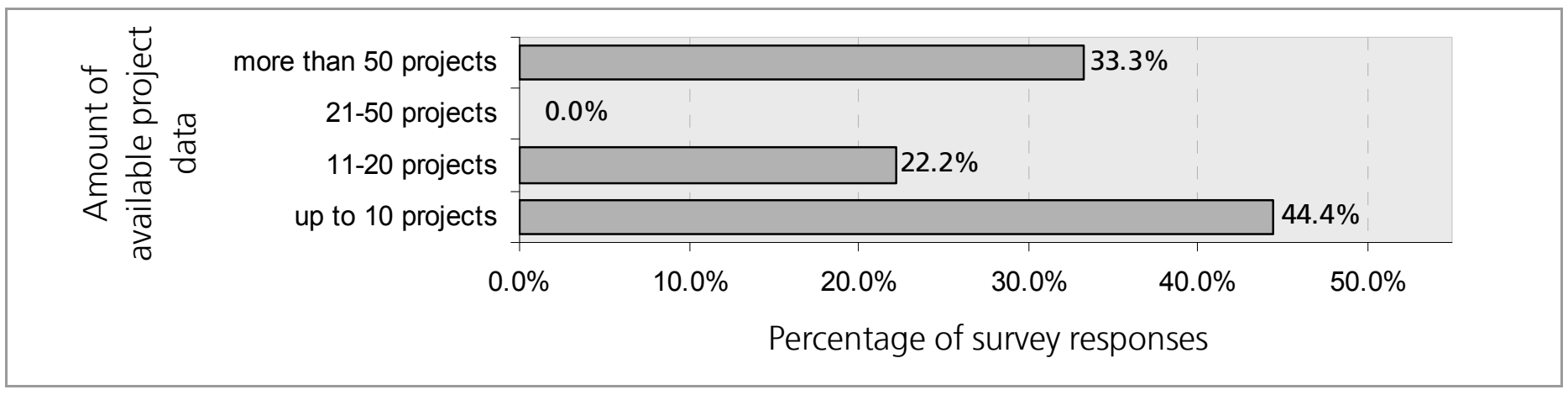

**Fig. 4.** Historical project data available in the software industry

One of the factors responsible for insufficient data availability [11, 13] is the rapid technological advancement in the software engineering domain. Software organizations focusing on large projects, such as those in the space domain, might never be able to collect a sufficient amount of up-to-date data.

## 4.3 Applied Effort Estimation Methods

Quite surprising in the light of available manpower seems to be the observation that a vast majority of software organizations employ effort estimation based on expert assessment (Fig. 5). The industrial surveys showed that 9 out of 10 surveyed companies employed estimation based on human judgment. This remains in line with observations made for over a decade, where about 60% to 85% of software projects rely exclusively upon expert estimates [14, 16]. Ad-hoc estimates are, however, rare and predictions are typically obtained through a group meeting where several experts follow a structured estimation process (such as that implemented in the Delphi method [1]) to come up with a final effort prediction. One of the major reasons is distrust of data-driven methods, which is the result of the lack of substantial evidence in favor of those methods [14]. Moreover, they are often perceived as complicated to use, and thus requiring significant overhead. Therefore, if used at all, only simple data-driven methods such as linear regression based on the effort and size of already finished projects are applied - often in combination with expert judgment [15].

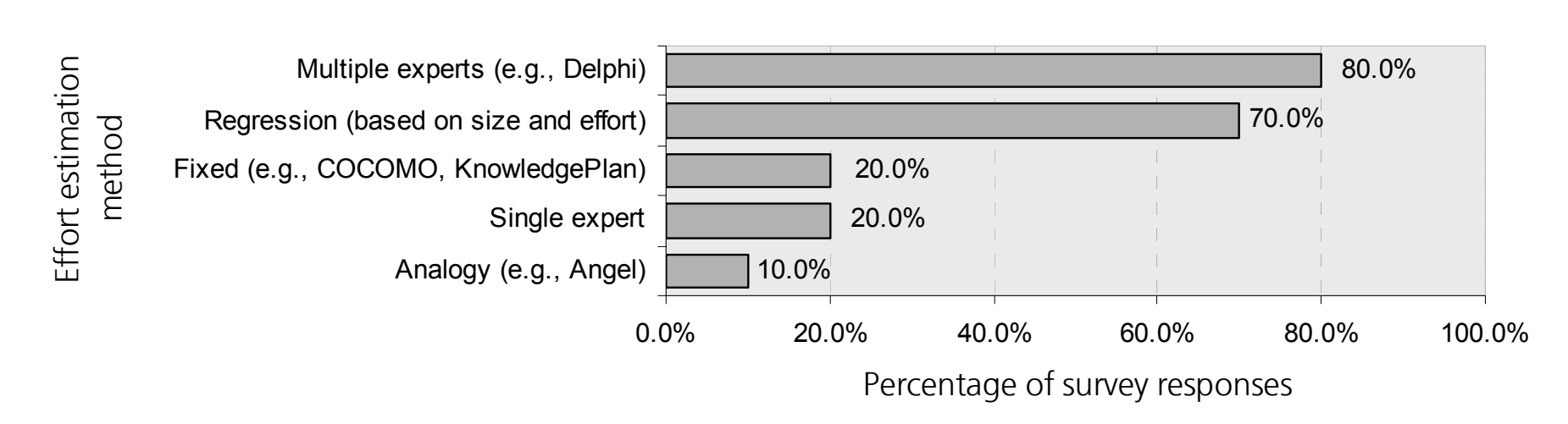

**Fig. 5.** Effort estimation methods applied in the software industry

Experts estimate software size (usually through a work breakdown structure), for instance, in combination with the average development productivity of already completed projects to come up with the effort prediction (*Effort = Size / AverageProductivity*). A simple combination of the individual results of expert- and data-based estimates typically does not seem to provide significant improvement compared to expert estimates alone [15]. Finally, the use of more sophisticated methods, such as COCOMO [1] or decision trees [3], remains marginal.

## 4.4 Detailed Requirements regarding Software Effort Estimation Methods

An investigation of industrial objectives and abilities regarding effort estimation as well as an analysis of detailed comments given by study respondents led us to define a detailed list of requirements that should be considered when selecting a particular estimation method:

1. *Expert involvement*: The method does not require extensive expert involvement, i.e., it requires a minimal number of experts, limited involvement (effort) per expert, and minimal expertise.
2. *Required data*: The method does not require large amounts of measurement data of a specific type (i.e., measurement scale) and distribution (e.g., normal).
3. *Robustness*: The method is robust to low-quality data inputs, i.e., incomplete (e.g., missing data), inconsistent (e.g., data outliers), redundant, and collinear data.
4. *Flexibility*: The method is free from a specific estimation model and provides context-specific outputs.

5. *Complexity*: The method has limited complexity, i.e., it does not employ many techniques, its underlying theory is easy to understand, and it does not require specifying many sophisticated parameters.
6. *Support level*: There is comprehensive support provided along with the method, i.e., complete and understandable documentation, and a useful software tool.
7. *Handling uncertainty*: The method supports handling the uncertainty of the estimation (i.e., inputs and outputs).
8. *Comprehensiveness*: The method can be applied for estimating different kinds of project activities (e.g., management, engineering) on various levels of granularity (e.g., project, phase, and task).
9. *Availability*: The method can be applied during all stages (phases) of the software development lifecycle.
10. *Empirical evidence*: There is comprehensive empirical evidence supporting the theoretical and practical validity of the method.
11. *Informative power*: The method provides complete and understandable information that supports the achievement of numerous estimation objectives (e.g., effective effort management). In particular, it provides context-specific information regarding relevant effort factors, their interactions, and their impact on effort.
12. *Reliability*: The method provides the output that reflects the true situation in a given context. In particular, it provides accurate, precise, and repeatable outputs.
13. *Portability*: The method provides estimation outputs that are either applicable in other contexts without any modification or are easily adaptable to other contexts.

These requirements may be further quantified, e.g., using a Likert-scale with respect to their value and importance, and used within multi-criteria decision support [17] for selecting the best suitable effort estimation method within a specific project context.

## 5 Overview of Existing Software Effort Estimation Methods

Existing effort estimation methods basically differ with respect to the type of inputs they require and the form of the estimation model they provide. With respect to input data, we differentiate between three major groups: data-intensive, expert-based, and hybrid methods (combining available data and expert knowledge in order to come up with estimates). An analysis of existing estimation methods with respect to industrial objectives and derived requirements indicates a few leading methods that meet most of the requirements; although no single method satisfies all requirements [20].

The major point against *data-intensive* methods is that they require large data sets. This is not the typical industrial situation (even in high-maturity, process-oriented organizations), which is rather characterized by sparse, incomplete, and inconsistent data. Moreover, these methods are often complicated to use, and have not actually proven to be superior to expert-based methods. Among the *data-intensive methods*, some require past project data for building customized models (*define-your-own-model approaches*), others provide an already defined model, where factors and their relationships are fixed based on a set of multi-organizational project data (*fixed-model approaches*). The major advantage of fixed-model approaches is that they, theoretically, do not require any historical data to be applied. Those methods might be especially attractive in the IV&V context, where very sparse (if any) data are typically available. Yet, in practice, fixed models, such as COCOMO [1], are developed for a specific context and are, by definition, only suited for estimating the types of projects for which the fixed model was built. The applicability of such models for the IV&V context is, in practice, very limited. In order to improve their performance, a significant amount of organization-specific project data would be required for calibrating the generic model. In that case, the usefulness of the fixed-model approaches for IV&V effort estimation would not differ much from the define-your-own-model approaches, which require a significant amount of reliable, context-specific data to build customized effort models. Application of the define-your-own-model methods is further limited by the additional requirements of specific methods. Parametric approaches, such as regression [18], for instance, make several assumptions about underlying project data (completeness, normal distribution, etc.) that are rarely met in the software domain. Non-parametric methods originating from the machine learning domain, such as artificial neural networks (ANN) [2] or Decision Trees [3], make practically no assumptions about the data but are quite sensitive to their parameter configuration, and there is usually little universal guidance regarding how to set those parameters. Thus, finding appropriate parameter values requires some preliminary experimentation.

In contrast to data-intensive methods, *expert-based* estimation does not require any project measurement data. However, it is widely criticized due to large overheads and the requirement for seasoned experts each time the estimation needs to be performed. Moreover, the reliability of the outputs it provides largely depends on the expertise and individual preferences of the human experts involved. Moreover, since the rationale underlying the final estimates is not modeled explicitly, expert-based estimation, by itself, provides hardly any support for effective decision making (risk management, process improvement, negotiations, etc.). Even though experts identify the factors influencing development effort, we find that they typically tend to largely disagree on them and omit relevant factors while selecting irrelevant ones.

Recently, a few *hybrid* methods have been proposed to cope with the deficits of data-intensive and expert-based estimation. One of the main objectives of hybrid estimation is to reduce the amount of both measurement data and human expertise by combining those two sources of information. In consequence, more reliable estimates should be obtained with reduced overhead. Empirical applications [12, 21] report on higher estimation accuracy and stability of hybrid methods when compared to those based solely on data or experts. Moreover, methods that employ explicit causal effort modeling [12, 21] have proven to greatly contribute to the achievement of a variety of organizational objectives, such as risk management or process/productivity improvement. Yet, the causal effort model, from which an effort model is derived, is typically developed based either solely on experts or on data. This still requires either significant involvement of experienced human experts or large amounts of high-quality data. Moreover, the reliability of a causal model based on homogeneous sources of information is typically limited. We found, for example, that domain experts, driven by their subjective preferences, tend to omit relevant causal effects (i.e., effort drivers and their causal interactions) while choosing irrelevant ones.

# 6 Summary and Further Work Directions

In this paper, we analyzed industrial trends with respect to software effort estimation. In particular, we were interested in what the objectives of effort estimation are, what the industrial capabilities for applying certain estimation strategies are, and finally, which of the existing methods are actually employed within the software industry.

To summarize our findings, there is a growing need for supporting various project and process management activities, such as risk management, project negotiations, or process improvement. Effort estimation methods that grew upon traditional planning objectives usually focus on providing an accurate point estimate without giving much insight into organization-specific causal effort dependencies, in particular into the most relevant factors contributing to project effort. In consequence, even though accurate estimates can be provided, there is hardly any support for decision making in situations where the estimate exceeds the available budget. Moreover, software organizations, even high-maturity ones, do not have sufficient resources to apply existing estimation methods, which typically require either extensive involvement of seasoned domain experts or large amounts of high-quality measurement data. In consequence, estimation performed with insufficient resources (manpower) by multiple human experts is still a common industrial practice. Even though quantitative methods are applied, they are rather simple and often based on sparse and unreliable data. In this situation, hybrid methods that combine data analysis with expert judgment and provide a transparent, context-specific model of causal effort dependencies seem to offer a potential remedy. Inclusion of subjective elements, such as an expert's opinion, in analytical models potentially allows for a significant reduction in the number of irrelevant variables considered in a model, as well as accounting for factors that are difficult to measure. Yet, the few hybrid methods that have been proposed so far base the development of an explicit causal effort model either on measurement data or on human judgment. In consequence, they inherit the major weaknesses of data-driven and expert-based estimation.

The discrepancy between what is needed by the software industry and what is actually provided by the research community indicates that further research should, in general, focus on providing estimation methods that keep up with current industrial needs and abilities. In particular, methods that integrate analysis of sparse data with minimal involvement of human experts in order to come up with causal effort models should be investigated.

# References

1. Boehm, B.W., Abts, C., Brown, A.W., Chulani, S., Clark, B.K., Horowitz, E., Madachy, R., Refer, D., Steece B.: Software Cost Estimation with COCOMO II. Prentice Hall (2000)
2. Boetticher, G.: An Assessment of Metric Contribution in the Construction of a Neural Network-Based Effort Estimator. In: International Workshop Soft Computing Applied to Software Engineering, pp. 59–65 (2001)
3. Breiman, L., Friedman, J., Ohlsen, R., Stone, C.: Classification and Regression Trees. Wadsworth & Brooks/Cole Advanced Books & Software (1984)

4. Brereton, P., Kitchenham, B.A., Budgen, D., Turner, M., Khalil, M.: Lessons from Applying the Systematic Literature Review Process within the Software Engineering Domain. Journal of Systems and Software, vol. 80, pp. 571–583 (2007)
5. Briand, L.C., Wieczorek, I.: Resource Modeling in Software Engineering. In: Marciniak, J.J. (ed.): Encyclopedia of Software Engineering. 2nd edn. Wiley (2002)
6. Charette, R.N.: Why Software Fails [Software Failure]. IEEE Spectrum, vol. 32, no. 9, pp. 42–49 (2005)
7. Jørgensen, M., Løvstad, N., Moen L.: Combining Quantitative Software Development Cost Estimation Precision Data with Qualitative Data from Project Experience Reports at Ericsson Design Center in Norway. In: International Conference on Empirical Assessments of Software Engineering (2002)
8. Jørgensen, M., Shepperd, M.: A Systematic Review of Software Development Cost Estimation Studies. IEEE Transactions on Software Engineering, vol. 33, no. 1, pp. 33–53 (2007)
9. Kitchenham, B.: Procedures for Performing Systematic Reviews. Technical report TR/SE0401, Software Engineering Group, Keele University (2004)
10. Kläs, M., Trendowicz, A., Wickenkamp, A., Münch, J., Kikuchi, N., Ishigai, Y.: The Use of Simulation Techniques for Hybrid Software Cost Estimation and Risk Analysis. Advances in Computers, vol. 74, pp. 115–174 (2008)
11. MacDonell, S.G., Shepperd, M.J.: Comparing Local and Global Software Effort Estimation Models – Reflections on a Systematic Review. In: International Symposium on Empirical Software Engineering & Measurement, pp. 401–409 (2007)
12. Mendes E.: A Comparison of Techniques for Web Effort Estimation. In: International Symposium on Empirical Software Engineering and Measurement, pp. 334–343 (2007)
13. Mendes, E., Lokan, C.: Replicating Studies on Cross- vs Single-company Effort Models Using the ISBSG Database. Journal of Empirical Software Engineering, vol. 13, no. 1, pp. 3–37 (2008)
14. Moløkken-Østvold, K.J., Jørgensen, M.: Expert Estimation of the Effort of Web-Development Projects: Why Are Software Professionals in Technical Roles More Optimistic Than Those in Non-Technical Roles? Journal of Empirical Software Engineering, vol. 10, no. 1, pp. 7–29 (2005)
15. Moløkken-Østvold, K.J., Jørgensen, M., Tanilkan, S.S., Gallis, H., Lien, A.C., Hove, S.E.: A Survey on Software Estimation in the Norwegian Industry. In: International Symposium on Software Metrics, pp. 208–219 (2004)
16. Moløkken-Østvold, K.J., Jørgensen, M.: A Review of Surveys on Software Effort Estimation. In: International Symposium on Empirical Software Engineering, pp. 223–230 (2003)
17. Paschetta, E., Andolfi, M., Costamanga, M., Rosenga, G.: A Multicriteria-based Methodology for the Evaluation of Software Cost Estimation Models and Tools. In: International Conference on Software Measurement and Management (1995)
18. Sentas, P., Angelis, L., Stamelos, I., Bleris G.L: Software Productivity and Effort Prediction with Ordinal Regression. Journal of Information & Software Technology, vol. 47, no. 1, pp. 17–29 (2005)
19. The Standish Group: CHAOS Chronicles. Technical report, The Standish Group International, Inc. (2007)
20. Trendowicz, A.: Software Effort Estimation - Overview of Current Industrial Practices and Existing Methods. Technical report 06.08/E, Fraunhofer IESE, Kaiserslautern, Germany (2008)
21. Trendowicz, A., Heidrich, J., Münch, J., Ishigai, Y., Yokoyama, K., Kikuchi, N.: Development of a Hybrid Cost Estimation Model in an Iterative Manner. In: International Conference on Software Engineering, pp. 331–340 (2006)